\documentclass[aps,prb,notitlepage,floatfix,twocolumn,superscriptaddress]{revtex4-1}
\usepackage{graphicx}
\usepackage{amsfonts}
\usepackage[figuresright]{rotating}
\usepackage{amssymb}
\usepackage{amsmath}
\usepackage{braket}
\usepackage{booktabs}
\usepackage{slashed}
\usepackage[OT1]{fontenc}
\usepackage{verbatim}
\usepackage{txfonts}
\usepackage[colorlinks,linkcolor=blue,anchorcolor=blue,citecolor=blue,urlcolor=blue]{hyperref}
\usepackage[dvipsnames]{xcolor}
\usepackage[titletoc]{appendix}

\usepackage{tikz,fp}
\usepackage{tikz-cd}
\usetikzlibrary{arrows}
\usetikzlibrary{intersections}
\usetikzlibrary{shapes.geometric}
\usetikzlibrary{decorations.pathmorphing, patterns,shapes,fixedpointarithmetic}
\usetikzlibrary{decorations.markings}

\pgfdeclarepatternformonly{south west lines}{\pgfqpoint{-0pt}{-0pt}}{\pgfqpoint{3pt}{3pt}}{\pgfqpoint{3pt}{3pt}}{
	\pgfsetlinewidth{0.4pt}
	\pgfpathmoveto{\pgfqpoint{0pt}{0pt}}
	\pgfpathlineto{\pgfqpoint{3pt}{3pt}}
	\pgfpathmoveto{\pgfqpoint{2.8pt}{-.2pt}}
	\pgfpathlineto{\pgfqpoint{3.2pt}{.2pt}}
	\pgfpathmoveto{\pgfqpoint{-.2pt}{2.8pt}}
	\pgfpathlineto{\pgfqpoint{.2pt}{3.2pt}}
	\pgfusepath{stroke}}

\tikzset{
	mid arrow/.style={postaction={decorate,decoration={
				markings,
				mark=at position .575 with {\arrow{stealth}}
	}}},
	near arrow/.style={postaction={decorate,decoration={
				markings,
				mark=at position .275 with {\arrow{stealth}}
	}}},
	far arrow/.style={postaction={decorate,decoration={
				markings,
				mark=at position .800 with {\arrow{stealth}}
	}}},
	snake arrow/.style={fixed point arithmetic, decorate, decoration={snake,amplitude=2pt, segment length=11pt},postaction={decoration={markings,mark=at position 0.625 with {\arrow{stealth}}},decorate}},
}

\newcommand{\SL}{\operatorname{SL}}

\newcommand{\iu}{{i\mkern1mu}}

\newcommand{\RN}[1]{%
	\textup{\uppercase\expandafter{\romannumeral#1}}%
}

\begin{document}
	
	\title{Tunneling through an Eternal Traversable Wormhole}
	\author{Tian-Gang Zhou}
	\affiliation{Institute for Advanced Study, Tsinghua University, Beijing,100084, China}
	
	\author{Pengfei Zhang}
	\thanks{PengfeiZhang.physics@gmail.com}
	\affiliation{Institute for Quantum Information and Matter and Walter Burke Institute for Theoretical Physics, California Institute of Technology, Pasadena, California 91125, USA}
	\date{\today}
	
	\begin{abstract}
		The Maldacena-Qi model describes two copies of the Sachdev-Ye-Kitaev model coupled with an additional coupling, and is dual to the Jackiw-Teitelboim gravity which exhibits an eternal traversable wormhole in the low-temperature limit. In this work, we study an experimental consequence of the existence of the traversable wormhole by considering the tunneling spectroscopy for the Maldacena-Qi model. Comparing to the high-temperature black hole phase where the bulk geometry is disconnected, we find that both the tunneling probability and the differential conductance in the low-temperature wormhole phase show non-trivial oscillation, which directly provides an unambiguous signature of the underlying $\operatorname{SL}(2)$ symmetry of the bulk geometry. We also perform bulk calculations in both high and low-temperature phases, which match the results from the boundary quantum theory.
	\end{abstract}
	
	\maketitle

	\section{Introduction} Holographic duality has established many valuable connections between some strongly interacting quantum systems and the semi-classical gravity theory \cite{hartnoll2018holographic}. The Sachdev-Ye-Kitaev (SYK) model \cite{kitaev2014hidden,sachdev1993gapless,maldacena2016remarks,kitaev2018soft}, which describes $N$ randomly interacting Majorana zero modes, is one of the concrete examples where the holographic description exists. In the large-$N$ and low-temperature limit, the model effectively describes the Jackiw-Teitelboim gravity in two-dimensional nearly anti-de Sitter (AdS$_2$) spacetime \cite{maldacena2016conformal}. Later, several generalizations of the SYK model have been introduced to study different physics \cite{Gnezdilov_2018,Kruchkov_2020,Altland_2019,can2019charge,sachdev2015bekenstein,davison2017thermoelectric,gu2020notes,maldacena2018eternal,banerjee2017solvable,chen2017competition,zhang2017dispersive,jian2017solvable,gu2017local,khveshchenko2019one,khveshchenko2020connecting,klebanov2020spontaneous}, including the tunneling spectroscopy \cite{Gnezdilov_2018,Kruchkov_2020,Altland_2019,can2019charge} for generalizations with $U(1)$ symmetry \cite{sachdev2015bekenstein,davison2017thermoelectric,gu2020notes}. Moreover, the quantum simulation of the SYK model \cite{garcia2017digital} has been performed in NMR systems \cite{Luo_2019} and there are several other proposals for realizing the SYK model in different experimental systems \cite{danshita2017creating,Chen_2018,wei2020optical,chew2017approximating}. 
	
	Meanwhile, wormholes have become a central topic in the fields of gravity and quantum information. It is an essential entry point to understand the quantum teleportation \cite{gao2017traversable,maldacena2017diving}, the late time behavior of the spectral form factor \cite{saad2018semiclassical}, and the resolution of the information paradox \cite{almheiri2019replica,penington2019replica}. Besides, a simple quantum model for wormholes is the coupled SYK model introduced by Maldacena and Qi \cite{maldacena2018eternal}. The Maldacena-Qi (MQ) model consists of two copies (left/right) of the original SYK model, with additional direct hopping between corresponding modes. Each copy of the SYK model corresponds to a boundary of the AdS spacetime. In the low-temperature limit, a traversable wormhole \cite{gao2017traversable,maldacena2017diving} between the left and the right boundary is formed. At higher temperatures, there is a first-order transition to a geometry with two disconnected black holes. Dynamical evolution and equilibrium properties of the MQ model have been studied in \cite{Plugge_2020,Qi_2020,maldacena2019syk}. Later, the model is generalized into the complex fermion version with $U(1)$ symmetry \cite{sahoo2020traversable,sorokhaibam2020traversable} and is found to be related to the large-$M$ random spin models \cite{zhou2020disconnecting}. There is also an experimental proposal for realizing the MQ model \cite{lantagne2020diagnosing}.
	
\begin{figure}[tb]
		\centering
		\includegraphics[width=0.9\linewidth]{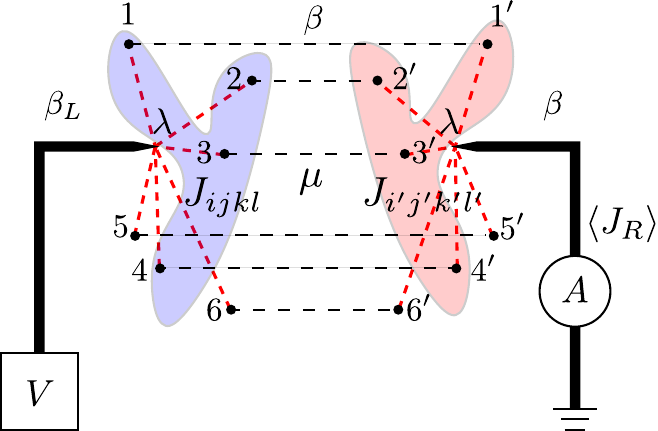}
		\caption{Schematics of the setup where we couple each side of the MQ model to a lead that allows to measure the tunneling current. Here the blue/red blob represents specific SYK interaction terms for the $L/R$ copy, which acts non-trivialy on four fermion modes. $V$ represents bias voltage added to the left lead and the current on the right lead $J_R$ is measured by the ammeter $A$. The inverse temperature is $\beta_L$ for the left lead, $\beta$ for the right lead, and $\beta$ for the complex MQ model system.}
		\label{fig:wormholeleads}
	\end{figure}

	In this work, we study the experimental consequence of the eternal traversable wormhole in the bulk from the transport perspective. As in the conventional experimental setup for measuring the tunneling current, we consider attaching leads to each side of the complex MQ model, as shown in Fig.~\ref{fig:wormholeleads}. We then apply a voltage at the left lead and measure the current through the right lead. Intuitively, when the MQ model is in the wormhole phase, an electron in the left lead dives into the traversable wormhole and will appear in the right lead, leading to large tunneling probability when the energy matches the intrinsic modes of the AdS$_2$ spacetime. On the other hand, if the MQ model is in the black hole phase, the correlation between the two sides becomes much weaker, and the tunneling probability becomes small. We will show that this intuition is indeed consistent with detailed calculations on either the quantum side or the gravity side, and the tunneling spectroscopy provides an unambiguous signature of the bulk geometry.

	\section{Model}
	We consider coupling each side of the complex MQ model to a different lead described by non-interacting fermions. The Hamiltonian is
	\begin{equation}
	\hat{H}=\hat{H}_{\text{MQ}}+\hat{H}_{\text{Lead}}+\sum_{i,p}\left(\lambda_i \hat{c}^\dagger_{L,i}\hat{\psi}^{}_{L,p}+\lambda_i\hat{c}^\dagger_{R,i}\hat{\psi}^{}_{R,p}+\text{H.C.}\right),
	\end{equation}
	where $\hat{H}_{\text{MQ}}$ and $\hat{H}_{\text{Lead}}$ read
	\begin{equation}
	\begin{aligned}
	\hat{H}_{\text{MQ}}=&\sum_{i<j;k<l}J_{ij,kl}\left(\hat{c}^{\dagger}_{L,i}\hat{c}^{\dagger}_{L,j}\hat{c}^{}_{L,k}\hat{c}^{}_{L,l}+\hat{c}^{\dagger}_{R,i}\hat{c}^{\dagger}_{R,j}\hat{c}^{}_{R,k}\hat{c}^{}_{R,l}\right)\\
	&+\mu\sum_i\left(\hat{c}^{\dagger}_{L,i}\hat{c}^{}_{R,i}+\hat{c}^{\dagger}_{R,i}\hat{c}^{}_{L,i}\right),\\
	\hat{H}_{\text{Lead}}=&\sum_{p}\epsilon_p\left(\hat{\psi}^{\dagger}_{L,p}\hat{\psi}^{}_{L,p}+\hat{\psi}^{\dagger}_{R,p}\hat{\psi}^{}_{R,p}  \right).
	\end{aligned}
	\end{equation}
	Here $i,j,k,l=1,2...N$. $\hat{c}^{}_{L/R,i}$ is the annihilation operator in the left/right copy of the SYK model, where two copies are coupled by $\mu$. $\hat{\psi}^{}_{L/R,p}$ describes fermions in the left/right lead with momentum $p$ with a gapless dispersion $\epsilon_p$. Note that the number of modes in leads does not scales with $N$. $J_{ij,kl}$, and $\lambda_i$ are independent random Gaussian variables with zero mean and variance 
	\begin{equation}
	\overline{|J_{ij,kl}|^2}=\frac{2J^2}{N^3},\ \ \ \ \ \ \overline{|\lambda_i|^2}=\frac{\lambda^2}{N D}.
	\end{equation}
	Here $D$ is the number of coupled modes in the lead.
	
	We first consider the total system in thermal equilibrium at inverse temperature $\beta$. Since the leads only contain $O(1)$ modes, to the leading order of $1/N$, the two-point function of the MQ model is not modified by a finite $\lambda$. Consequently, the system still contains two different phases. In the low-temperature limit, the MQ model is dual to an eternal traversable wormhole, where two copies of the SYK model are connected through a holographic bulk. Here the perfect correlation between two copies plays an important role for obtaining a wormhole solution in the low-temperature limit \cite{maldacena2018eternal}. At higher temperatures, the system turns into a black holes phase where the emergent spacetimes in the gravitational description correspond to nearly disconnected black holes.
	
	We then analyze modes in the leads. Without the coupling to the SYK dots, the left and the right leads are decoupled. When $\lambda$ is turned on, the fermion can tunnel from the left lead to the right lead through the MQ model. Explicitly, we define the retarded Green's function $G^\mathcal{R}_{O}(t)\equiv -i\theta(t)\left<\left[\hat{O}^{}(t),\hat{O}^\dagger(0)\right]\right>,$ and focus on the local fermion modes $\hat{\psi}_{\alpha}=\sum_p\hat{\psi}^{}_{\alpha,p}/\sqrt{D}$ with $\alpha =L/R$ that couple to the MQ model. In other words, we put the contact point of the lead to the SYK models at $x = 0$.
	Taking into account the self-energy from the MQ model, the Schwinger-Dyson equation of $\psi$ reads
	\begin{equation} \label{SDR}
	\left(G^\mathcal{R}_{\psi}(\omega)^{-1}\right)_{\alpha\gamma}=\left(G^{\mathcal{R},0}_{\psi }(\omega)^{-1}\right)_{\alpha\gamma}-\lambda^2G^\mathcal{R}_c(\omega)_{\alpha\gamma}.
	\end{equation}
	Here $G^\mathcal{R}_c(\omega)_{\alpha\gamma}$ is the retarded Green's function of the MQ model ($\alpha\gamma$ component). $G^{\mathcal{R},0}_{\psi }(\omega)_{\alpha\gamma}=-i\pi\rho_0 \delta_{\alpha\gamma}$ is the bare Green's function of leads. Here we assume that the density of states of leads can be approximated as a constant $\rho_0$. Other real-time Green's functions $G^\mathcal{A}_{\psi}$, $G^\mathcal{K}_{\psi}$ are then determined by $G^\mathcal{A}_{\psi}(\omega)=G^\mathcal{R}_{\psi}(\omega)^\dagger$ and at inverse temperature $\beta$ the fluctuation-dissipation theorem gives:
	\begin{equation}\label{FDT}
	G^\mathcal{K}_{\psi}(\omega)_{\text{eq}}=\left(G^\mathcal{R}_{\psi}(\omega)-G^\mathcal{A}_{\psi}(\omega)\right)F_\beta(\omega),
	\end{equation}
	where $F_\beta(\omega)=1-2f^F_\beta(\omega)=\tanh(\beta\omega/2)$, with $f^F_\beta(\omega)$ being the Fermi-Dirac distribution function.

	\section{Tunneling Spectroscopy} Now we consider the non-equilibrium problem of calculating the tunneling current. The charge of the right lead is defined as $\hat{Q}_R=\sum_p\hat{\psi}^{\dagger}_{R,p}\hat{\psi}^{}_{R,p}$. Consequently, the current flowing from the right SYK model to the right lead reads
	\begin{equation}
	\begin{aligned}
	\hat{J}_R&=-i[\hat{H},\hat{Q}_R]=-i\sum_{i,p}(\lambda_i\hat{c}^\dagger_{R,i}\hat{\psi}^{}_{R,p}-\lambda_{i}^*\hat{\psi}^\dagger_{R,p}\hat{c}^{}_{R,i}).
	\end{aligned}
	\end{equation}
	
	When the full system is in thermal equilibrium, the current vanishes. We are interested in the setup shown in Fig.~\ref{fig:wormholeleads}, where we tune the voltage $V$ and the inverse temperature $\beta_L$ of the left lead while fixing all other parts of the system to $V=0$ and $\beta$. Measuring $\left< \hat{J}_R \right>$ then detects how many particles from the left lead can tunnel through the MQ model.
	
	The calculation of tunneling current can be analyzed using the Schwinger-Keldysh formalism. The contours of such path-integral method contain a forward and a backward evolution with fermionic field $c_{\alpha,i,f/b}$ and $\psi_{\alpha,p,f/b}$. After the Keldysh rotation 
	$$c_{\alpha,i,1\,/\,2}=c_{\alpha,i,f}\pm c_{\alpha,i,b},\ \ \ \ \ \ \overline{c}_{\alpha,i,1\,/\,2}=\overline{c}_{\alpha,i,f}\mp \overline{c}_{\alpha,i,b},$$ and similarly for $\psi_{\alpha,p,1\,/\,2}$. The Green's function $\mathsf{G}_O(\omega)_{\alpha\gamma}^{ab}$ becomes $4\times 4$ matrices with both $\alpha,\gamma=L/R$ and $a,b=1\,/\,2$ label. Explicitly, in $1\,/\,2$ space we have
	\begin{equation}
	\mathsf{G}_\psi(\omega)_{\alpha\gamma}=
	\begin{pmatrix}
	G^\mathcal{R}_{\psi}(\omega)_{\alpha\gamma}&G^\mathcal{K}_{\psi}(\omega)_{\alpha\gamma}\\
	0&G^\mathcal{A}_{\psi}(\omega)_{\alpha\gamma}
	\end{pmatrix},
	\end{equation}
	and similarly for $\mathsf{G}_c(\omega)_{\alpha\gamma}$. As in eq.~\eqref{SDR}, the Schwinger-Dyson equation for $\mathsf{G}_\psi(\omega)$ now reads:
	\begin{equation}
	\left(\mathsf{G}_\psi(\omega)^{-1}\right)^{ab}_{\alpha\gamma}=\left(\mathsf{G}^0_\psi(\omega)^{-1}\right)^{ab}_{\alpha\gamma}-\lambda^2\left(\mathsf{G}_c(\omega)\right)^{ab}_{\alpha\gamma}.
	\end{equation}
	Similar to the equilibrium case, $\mathsf{G}_c$ is the equilibrium real-time Green's function of MQ model, which satisfies the fluctuation-dissipation theorem eq.~\eqref{FDT} at $\beta$ without additional chemical potential. On the contrary, $\mathsf{G}_\psi$ does not satisfy the fluctuation-dissipation theorem since the left lead is at different temperature and chemical potential. Nevertheless, we have relations for the bare Green's functions of leads
	\begin{equation}
	\begin{aligned}
	G^{\mathcal{K},0}_{\psi}(\omega)_{LL}&=-2\pi i \rho_0F_{\beta_L}(\omega-V),\\
	G^{\mathcal{K},0}_{\psi}(\omega)_{RR}&=-2\pi i \rho_0F_{\beta}(\omega).
	\end{aligned}
	\end{equation}
	
	Then the tunneling current $\left<J_R\right>$ can be computed to the leading order under the $1/N$ expansion. Diagrammatically, we have
	\begin{equation}
	\begin{aligned}
	2\left<J_R\right>=&\begin{tikzpicture}[baseline={([yshift=0pt]current bounding box.center)}, scale=1.2]
	\draw[dashed,thick,mid arrow] (15pt,0pt)..controls (8pt,15pt) and (-8pt,15pt)..(-15pt,0pt);
	\draw[thick, mid arrow] (-15pt,0pt)..controls (-8pt,-15pt) and (8pt,-15pt)..(15pt,0pt);
	\filldraw  (15pt,0pt) circle (1pt);
	\node at (0pt,-18pt) {\scriptsize $i,\omega$};
	\node at (0pt,18pt) {\scriptsize $p,\omega$};
		\filldraw[white]  (-15pt,0pt) circle (2pt);
	\draw[thick]  (-15pt,0pt) circle (2pt);
	\end{tikzpicture}+
	\begin{tikzpicture}[baseline={([yshift=0pt]current bounding box.center)}, scale=1.2]
	\draw[dashed,thick,mid arrow] (15pt,0pt)..controls (11pt,8pt)..(8pt,12pt);
	\draw[dashed,thick,mid arrow] (-8pt,12pt)..controls (-11pt,8pt)..(-15pt,0pt);
	\draw[thick,mid arrow] (8pt,12pt)--(-8pt,12pt);
	\draw[thick, mid arrow] (-15pt,0pt)..controls (-8pt,-15pt) and (8pt,-15pt)..(15pt,0pt);
	\filldraw  (15pt,0pt) circle (1pt);
	\node at (0pt,-18pt) {\scriptsize $i,\omega$};
	\node at (0pt,18pt) {\scriptsize $j,\omega$};
	\node at (-17pt,12pt) {\scriptsize $p,\omega$};
	\node at (17pt,12pt) {\scriptsize $q,\omega$};
	\filldraw  (8pt,12pt) circle (1pt);
	\filldraw  (-8pt,12pt) circle (1pt);
	\filldraw[white]  (-15pt,0pt) circle (2pt);
	\draw[thick]  (-15pt,0pt) circle (2pt);
	\end{tikzpicture}+...
	\\&-(\text{Reverse all arrows}).
	\end{aligned}
	\end{equation}
	Here the dashed line represents $\psi$ fields and the solid line represents $c$ fields. The black dots represent the insertion of identity operator $\left( I \otimes I \right)$ in $L/R$ and $1\,/\,2$ space, while the open circles represent vertex $\lambda^2 \left( P_R\otimes\sigma_x \right)$. Here $P_R$ is the projector into the $R$ space and $\sigma_x$ is in the Pauli matrix in the $1\,/\,2$ space. Implicitly, all internal labels of modes, momentums and the frequency $\omega$ should be summed up or integrated over. Summing up all diagrams gives
	\begin{equation}
	\left<J_R\right>=\frac{\lambda^2}{2}\int \frac{d\omega}{2\pi}\text{tr}\left[\mathsf{G}_c \left( P_R\otimes\sigma_x \right)  \mathsf{G}_\psi
	-
	\mathsf{G}_\psi  \left(P_R\otimes\sigma_x\right)
	  \mathsf{G}_c\right].
	\end{equation}
	Here the trace is taken in both $L/R$ and $1\,/\,2$ space. Using the explicit formula for $\mathsf{G}_c$ and $\mathsf{G}_\psi$, we find 
	\begin{equation}\label{eq:current}
	\left<J_R\right>=\int \frac{d\omega}{2\pi}\left|T(\omega)\right|^2\left(f_{\beta_L}^F(\omega-V)-f_{\beta}^F(\omega)\right),
	\end{equation}
	with 
	\begin{equation}\label{eq:fullTomega2}
	|T(\omega)|^2=\left|\frac{2\Gamma(G^{\mathcal{R}}_c)_{LR}}{(\Gamma (G^{\mathcal{R}}_c)_{LL}-\Gamma(G^{\mathcal{R}}_c)_{LR}-i)(\Gamma (G^{\mathcal{R}}_c)_{LL}+\Gamma(G^{\mathcal{R}}_c)_{LR}-i)}\right|^2.
	\end{equation}
	Here we have defined $\Gamma=\pi\rho_0\lambda^2$. $|T(\omega)|^2$ can be understood as the tunneling probability from the left lead to the right lead. Explicitly, the tunneling current is zero if there is no left-right correlation in the MQ model. In the following sections, we analyze the tunneling current in different phases, focusing on the $\beta_L=\beta$ case. One can also derive a similar formula for the energy current with an additional factor of $\omega$ \cite{Kruchkov_2020}.
	
	\begin{figure}
		\centering
		\includegraphics[width=1.0\linewidth]{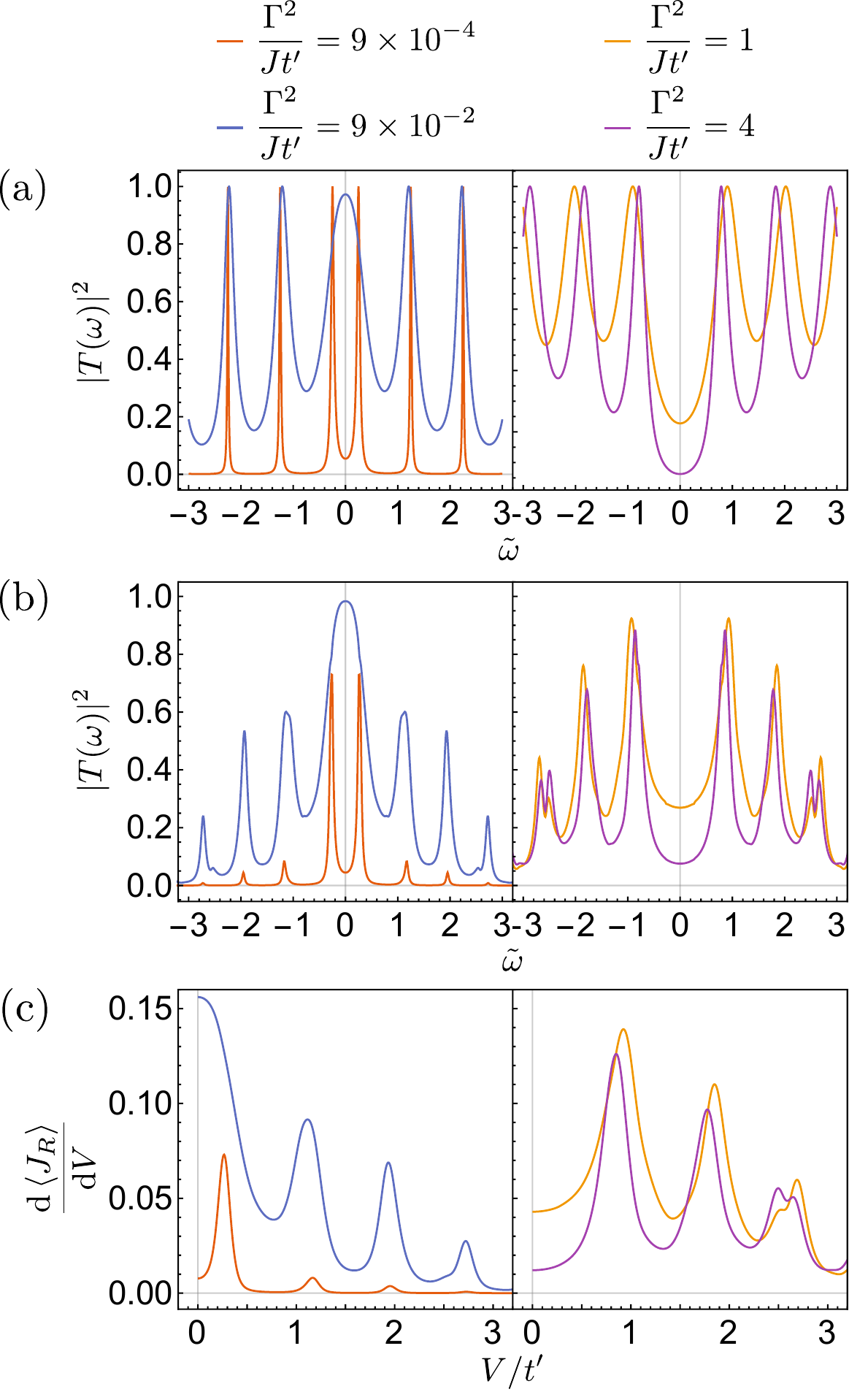}
		\caption{(a) The tunneling probability obtained from conformal solutions eq.~\eqref{wormhole res}. (b) The tunneling probability obtained by directly using the solutions of the Schwinger-Dyson equation in eq.~\eqref{eq:fullTomega2}. The Green's functions are self-consistently calculated in $\beta J=120$ and $\mu/J=0.025$, which results in $t'/J=0.3$ . (c) The differential conductance defined by taking the derivative of eq.~\eqref{eq:current}. We also use the same Green's functions as in (b). For each figure, left panel exhibits peaks at $\omega_n \equiv t'(n+1/4)$ and right panel exhibits peaks at $\omega'_n \equiv t'(n+3/4)$. }
		\label{fig:wormholenumerics}
	\end{figure}	

	\subsection{Wormhole Phase}
	We firstly consider the system in the zero-temperature limit and focus on small $\mu/J$. Holographically, the MQ model is dual to the eternal traversable wormhole geometry in the global AdS$_2$ spacetime with a metric $ds^2=(-dt_g^2+dx^2)/\sin^2 x$. The left/right copy of the SYK model lies on the boundary near $x=0$ or $x=\pi$. This indicates that both the diagonal and off-diagonal component of Green's functions are conformal \cite{maldacena2018eternal}. At zero temperature, after performing Fourier transform for the conformal Green's functions given in \cite{maldacena2018eternal}, we find
	\begin{equation}\label{Wormhole}
	\begin{aligned}
	G^\mathcal{R}_{c,\text{WH}}(\omega)_{\alpha\alpha}&=-\frac{2 \pi ^{5/4} \sin \left(\pi  \tilde{\omega}\right) \sec \left(2
		\pi  \tilde{\omega}\right)}{\sqrt{J t'} \mathcal{D}_{3/4}(\tilde{\omega})},\\
	G^\mathcal{R}_{c,\text{WH}}(\omega)_{\alpha\bar{\alpha}}&=-\frac{\mathcal{D}_{1/4}(\tilde{\omega})}{\sqrt{2} \pi ^{3/4} \sqrt{J
			t'}},
	\end{aligned}
	\end{equation}
	where $\alpha \neq \bar{\alpha}$. We have defined $\mathcal{D}_u(x)=\mathbf{\Gamma}(u-x)\mathbf{\Gamma}(u+x)$ and $\tilde{\omega}\equiv \omega/t'$ for conciseness. Here $\mathbf{\Gamma}(z)\equiv \int_0^\infty dx x^{z-1} e^{-x}$ is the standard gamma function. $t'$ is an additional parameter which is proportional to $J^{1/3}\mu^{2/3}$. It relates the global time $t_g$ to the boundary time $t$ as $t_g=t't$. The pole of $G^\mathcal{R}_c(\omega)$ lies at $|\omega|=\omega_n \equiv  t'(1/4+n) $ with $n=0,1,2...$ . This tower of the spectrum is fixed by the $\SL(2)$ symmetry of the AdS$_2$ spacetime. 
	
	The full expression for tunneling probability $|T(\omega)|^2$ can then be derived as: 
	\begin{equation}\label{wormhole res}
	|T(\omega)|^2=\frac{8}{\frac{4 \pi ^{5/2} \Gamma ^2}{J t'\mathcal{D}_{3/4}(\tilde{\omega})^2}+\frac{J t' \cos ^2\left(2 \pi  \tilde{\omega}\right) \mathcal{D}_{3/4}(\tilde{\omega})^2}{\pi ^{5/2} \Gamma ^2}-4
		\cos \left(2 \pi  \tilde{\omega}\right)+8}.
	\end{equation}
	  We plot $|T(\omega)|^2$ using the conformal solutions in Fig.~\ref{fig:wormholenumerics}(a). For extremely small coupling $\Gamma$, the fermion modes can only tunnel through the MQ model when they are on resonance with the MQ model. As a result, we have narrow peaks for the tunneling probability near $\omega_n$ with $|T(\omega)|^2\approx1$. This shows that the tunneling current is a direct probe of the eternal traversable wormhole and the underlying $\SL(2)$ symmetry. For larger coupling $\Gamma$, the narrow peaks get broadened, and the two peaks at $\pm \omega_0$ merge to a single broad peak. 
     
     If we further increase $\Gamma$, we find the peaks shift to $\omega_n'	 \equiv  \pm t'(3/4+n) $ at small $\omega$. At large $\Gamma$, eq.~\eqref{wormhole res} asymptotically approaches $(2 J t'\mathcal{D}_{3/4}(\tilde{\omega})^2)/(\pi ^{5/2} \Gamma ^2)$, which gives rise to the shift of peaks at $\omega_n'$.  Physically, this means that in the large $\Gamma$ limit, the low-energy tunneling is dominated by the contribution of bulk fields with scaling dimension $3/4$, which corresponds to the operator $\hat{\iota}_{\alpha,i}=\sum_{j;k<l}J_{ij,kl}\hat{c}^{\dagger}_{\alpha,j}\hat{c}^{}_{\alpha,k}\hat{c}^{}_{\alpha,l}$. Similar effects have been studied in the SYK chain model \cite{jian2018quantum} and give rise to the low-high voltage duality for the tunneling spectroscopy of the single SYK model \cite{Gnezdilov_2018}.


 We then consider the finite temperature and the finite $\mu/J$ corrections within the wormhole phase. In this case, we numerically solve the retarded Green's functions for the MQ model \cite{Plugge_2020,Qi_2020} and then compute the tunneling probability $|T(\omega)|^2$. As found in \cite{Plugge_2020,Qi_2020}, at finite temperature, the real-time Green's functions decay and there is a finite lifetime for quasi-particle modes. Moreover, the weight of modes decays quickly when $n$ increases. Consequently, as shown in Fig.~\ref{fig:wormholenumerics}(b), the tunneling probability, although still show a few peaks, becomes smaller than 1 when $n$ becomes large. We also plot differential conductance $d\langle J_R\rangle /dV$ in Fig.~\ref{fig:wormholenumerics}(c). As an essential physical quantity in condensed matter experiments, we find that differential conductance measurement at finite temperature also shows multiple peaks at $\omega_n$ or $\omega'_n$, providing the prominently non-trivial predictions for the experiments.

	\begin{figure}[tb]
		\centering
		\includegraphics[width=1.0\linewidth]{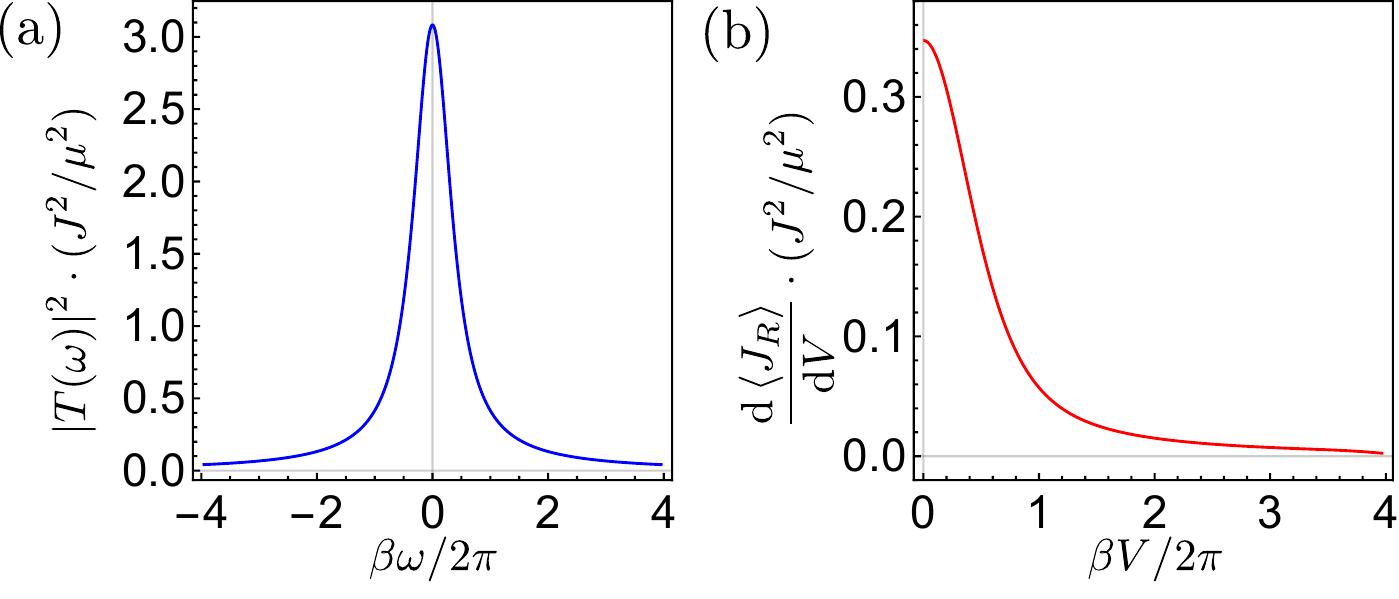}
		\caption{(a) The tunneling probability and (b) the differential conductance in the blackhole phase for $\Gamma/J=0.3$ and $\beta J=5$. }
		\label{fig:blackholenumerics}
	\end{figure}

	\subsection{Black Hole Phase} If we increase the temperature $T\gtrsim \mu$, there is a first-order transition in the MQ model after which the system gets into the black hole phase where each side lives on the boundary of a separate Rindler spacetime with $ds^2=-r(r-2)dt_R^2+dr^2/r(r-2)$. This is an analogy of the Hawking-Page transition in higher dimensions. In the black hole phase, from the gravity perspective, the Rindler spacetimes of the black holes emulated by left/right SYK models are disconnected. Adding the quantum correlation propotional to $\mu$, we can approximate
	\begin{equation}\label{SYKBH}
	G^\mathcal{R}_{c,\text{BH}}(\omega)_{\alpha\alpha}=-i\frac{ \sqrt{\beta/J }}{\sqrt{2}\pi ^{1/4} }\frac{\mathbf{\Gamma}\left(\frac{1}{4}-i\frac{\omega\beta}{2\pi}\right)}{\mathbf{\Gamma}\left(\frac{3}{4}-i\frac{\omega\beta}{2\pi}\right)},
	\end{equation}
	and $G^\mathcal{R}_{c,\text{BH}}(\omega)_{\alpha\bar{\alpha}}=\mu (G^\mathcal{R}_{c,\text{BH}}(\omega)_{\alpha\alpha})^2$. We have kept terms up to order $\mu$ and assume $\beta J \gg 1$. We can then expand $|T(\omega)|^2$ to the $O(\mu^2)$ order and obtain
	\begin{equation}\label{BH res}
	|T(\omega)|^2=\frac{4\mu^2/\Gamma^2}{\left|1+\sqrt{\frac{2J\pi^{1/2}}{\beta\Gamma^2}}\frac{\mathbf{\Gamma}\left(\frac{3}{4}-i\frac{\omega\beta}{2\pi}\right)}{\mathbf{\Gamma}\left(\frac{1}{4}-i\frac{\omega\beta}{2\pi}\right)}\right|^4}.
	\end{equation}
	Different from the wormhole phase, the spectral function is now a single peak near $\omega \sim 0$. We expect similar behavior for the tunneling probability, as verified in Fig.~\ref{fig:blackholenumerics}(a). We also plot differential conductance $d\left<J_R\right>/dV$ in Fig.~\ref{fig:blackholenumerics}(b), which, in contrast with the wormhole phase, shows no oscillating behavior. Plus, according to eq.~\eqref{BH res}, the magnitude of the tunneling probability is proportional to $(\mu/J)^2$ in the small $\mu/J$ limit, and is much smaller than the wormhole phase result in the order of magnitude 1.

	\section{Holographic Picture} 
	Now we analyze the problem from a bulk perspective. To have a simple holographic picture, we now choose a specific dispersion $\epsilon_p$. We assume that each lead can be described as a half-infinite line with both massless left-moving and right-moving Dirac fields living on a flat spacetime. we will show that a gravity calculation can reproduce previous results. 

	We first give a brief introduction of the holographic dictionary \cite{hartnoll2018holographic}. The holographic duality states that there is an equivalence between a strongly correlated quantum many-body system and a semi-classical system with gravity. The best understood example is the duality between an AdS gravity theory with spacetime dimension $d+1$ and a conformal field theory with spacetime dimension $d$, where the isometry group of the AdS spacetime matches the conformal group of the quantum theory. Mathematically, the holographic duality says
    \begin{equation}\label{duality}
    \int D\phi\ e^{iS_{\text{QFT}}[\phi]+\int JO}=\left.e^{iS_{\text{bulk}}[\Phi,g]}\right|_{\Phi\rightarrow J}.
    \end{equation}
	which is known as the GKPW formula. Here on the left hand side, we have a path integral for quantum fields $\phi$, with external source coupled $J$ to the field $O$. On the right hand side, we take the saddle point solution of a gravity theory, where we require that the boundary value of bulk fields should be equal to $J$, which will be specified in more details below. This formula enables us to compute the correlation function of a quantum system by solving classical equations.

\begin{figure}[tb]
		\centering
		\includegraphics[width=0.85\linewidth]{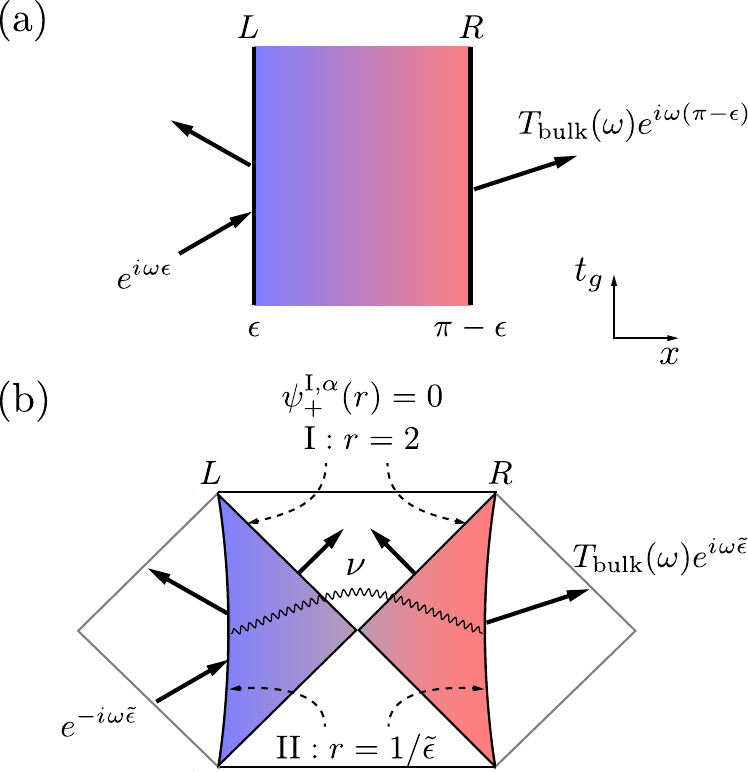}
		\caption{The gravity picture for the tunneling probability in the (a): wormhole geometry and (b): black hole geometry. Here the colored region in (a) corresponds to the AdS$_2$ spacetime, and the colored region in (b) corresponds to two copies of the Rindler spacetime. In both cases, we consider an in-going Dirac fermions from the left lead, and determine the tunneling amplitude $T_{\text{bulk}}(\omega)$ by solving the Dirac equation. For (b), the wiggle line corresponds to the coupling $\nu$ introduced in eq.~\eqref{cou}, and we impose the in-falling boundary condition $\psi_+^{\RN{1},\alpha}(r)=0$ at horizon $r=2$. }
		\label{fig:holographic}
	\end{figure}

	To illustrate how a holographic calculation works, we consider reproducing the Green's function eq.~\eqref{SYKBH} of the single SYK model from a bulk calculation. The calculation here is standard but may benefit general readers. The single SYK model in the low-energy limit is a conformal field theory in $0+1$ dimension, which, as mentioned in the last section, is dual to the AdS$_2$ Rindler geometry with metric $ds^2=-r(r-2)dt_R^2+dr^2/r(r-2)$. Such geometry can be illustrated as the left copy of the Fig.~\ref{fig:holographic}(b). This means one should imagine the bulk action contains some gravity part, with a solution of the corresponding metric. We then just need to solve the equation of motion for bulk fields on this geometry. Since eq.~\eqref{SYKBH} is the Green's function for fermionic operators, it is natural to consider the Dirac equation in the bulk:
	\begin{equation}
	S_{\text{bulk}}= \int \sqrt{-g}drdt_R\ \left(i\overline{\Psi}\slashed{D}\Psi - m\overline{\Psi}\Psi\right).
	\end{equation}
    Here we have included the mass $m$, which, as we will see, is related to the scaling dimension of the operator $\Delta$ in the quantum theory as $m=1/2-\Delta$. The Dirac equation in Rindler geometry can be solved exactly and the full expression can be found in the Appendix \ref{appendix}. Here we only focus on certain limits. 

    A general solution for the Dirac equation contains two independent coefficients $C_1$ and $C_2$, whose relation should be determined from boundary conditions. The spacetime contains two important limits---$\RN{1}$: the horizon at $r\rightarrow 2$ and $\RN{2}$: the spacetime boundary at $r\rightarrow \infty$. We firstly consider the solution near the horizon. Expanding the general solution \eqref{eq:S_DiracBH_full_solution} around $r=2$, one finds 
    \begin{equation}\label{eq:S_DiracBH_solution_req2}
		\begin{aligned}
				\psi_-^{\RN{1}}(r) =&(C_1+C_2) \frac{  2^{-\frac{1}{4}+\frac{i \omega }{2}} \pi  (r-2)^{-\frac{1}{4}-\frac{i
						\omega }{2}} \csc \left(\pi  \left(\frac{1}{2}+i \omega \right)\right)}{\mathbf{\Gamma} \left(\frac{1}{2}-i
				\omega \right) \mathbf{\Gamma} \left(i \omega +\frac{1}{2}\right)},  \\
				\psi_+^{\RN{1}}(r) =& 
				 \frac{  2^{-\frac{1}{4}-\frac{i \omega }{2}} \pi
					(r-2)^{-\frac{1}{4}+\frac{i \omega}{2} } \csc \left(\pi  \left(\frac{1}{2}-i \omega \right)\right) }{ \mathbf{\Gamma} \left(i
					\omega +\frac{1}{2}\right)^2} \\
				&\ \ \times\left(C_1 \frac{\mathbf{\Gamma}
					\left(i \omega +\frac{1}{2}+m\right)}{\mathbf{\Gamma} \left(\frac{1}{2}+m-i \omega \right)}-C_2\frac{\mathbf{\Gamma}
					\left(i \omega +\frac{1}{2}-m\right)}{\mathbf{\Gamma} \left(\frac{1}{2}-m-i \omega \right)}\right).
		\end{aligned}
	\end{equation}
    Here we use label $\RN{1}$ to represent the Dirac field near the horizon $r=2$, as we shown in Fig.~\ref{fig:holographic}(b). From the factor of $(r-2)^{\pm i\omega/2}$, we find that $\psi_-^{\RN{1}}$ represents a wave moving towards $r=2$, which is called an in-falling solution, and {$\psi_+^{\RN{1}}$} represents a wave moving outwards, which is called an out-going solution. Physically, one might expect a physical object should only contain in-falling components, which means we should impose the boundary condition that $\psi_+^{\RN{1}}=0$ when computing the retarded Green's function \cite{hartnoll2018holographic}. This determines the relation between $C_1$ and $C_2$:
	\begin{equation}\label{eq:S_DiracBH_solution_C1divC2}
		\begin{aligned}
			\frac{C_1}{C_2} = \frac{\mathbf{\Gamma} \left(\frac{1}{2}+m-i \omega \right) \mathbf{\Gamma} \left(i \omega +\frac{1}{2}-m\right)}{\mathbf{\Gamma}
				\left(\frac{1}{2}-m-i \omega \right) \mathbf{\Gamma} \left(i \omega +\frac{1}{2}+m\right)}.	
		\end{aligned}
	\end{equation}

	We then study the asymptotic behavior of the solution at the AdS boundary at $r\rightarrow \infty$. The result reads
    \begin{equation}\label{rinfBH}
    \begin{aligned}
     \begin{pmatrix}
     \psi_-^{\RN{2}}\\\psi_+^{\RN{2}}
     \end{pmatrix}=&C_2\frac{r^{-\frac{1}{2}+m}\mathbf{\Gamma}(1-m)\mathbf{\Gamma}(\frac{1}{2}-m+i\omega)}{2^m\mathbf{\Gamma}(1-2m)\mathbf{\Gamma}(\frac{1}{2}+i\omega)} \begin{pmatrix}
     1\\-1
     \end{pmatrix}\\
     +&C_1\frac{r^{-\frac{1}{2}-m}\mathbf{\Gamma}(1+m)\mathbf{\Gamma}(\frac{1}{2}+m+i\omega)}{2^{-m}\mathbf{\Gamma}(1+2m)\mathbf{\Gamma}(\frac{1}{2}+i\omega)} \begin{pmatrix}
     1\\1
     \end{pmatrix}.
     \end{aligned}
    \end{equation}
    Likewise, we use the label $\RN{2}$ for the result expanded near $r=\infty$. Now we should specify what is the boundary value of the bulk fields. Near $r\rightarrow \infty$, we see there are two independent solutions proportional to $r^{-1/2+m}(1,-1)$ and $r^{-1/2-m}(1,1)$. The boundary value of the bulk field is defined as the coefficient of the solution $r^{-1/2-m}(1,1)$. This relates the source $J(\omega)$ and $C_1$ 
    \begin{equation}
    J(\omega)=C_1\frac{\mathbf{\Gamma}(1+m)\mathbf{\Gamma}(\frac{1}{2}+m+i\omega)}{2^{-m}\mathbf{\Gamma}(1+2m)\mathbf{\Gamma}(\frac{1}{2}+i\omega)},
    \end{equation}
    which fixes the solution eq.~\eqref{eq:S_DiracBH_full_solution} in the bulk together with eq.~\eqref{eq:S_DiracBH_solution_C1divC2}. 

   One can then determine the right hand side of eq.~\eqref{duality} by using the solution, with additional boundary counter-terms for the holographic renormalization \cite{hartnoll2018holographic}. The result shows there is a shortcut to this calculation, where the coefficient of the other solution near $r\rightarrow \infty$ is just proportional to the expectation of boundary operator $\left<O(\omega)\right>_J$ with finite source term $J$:
    \begin{equation}
    \left<O(\omega)\right>_J \propto C_2\frac{\mathbf{\Gamma}(1-m)\mathbf{\Gamma}(\frac{1}{2}-m+i\omega)}{2^{m}\mathbf{\Gamma}(1-2m)\mathbf{\Gamma}(\frac{1}{2}+i\omega)}.
    \end{equation}
    The Green's function is then determined by 
   \begin{equation}
   G^R_{OO}(\omega)=\frac{\left<O(\omega)\right>_J}{J(\omega)}\propto\frac{4^m \mathbf{\Gamma} \left(m+\frac{1}{2}\right) \mathbf{\Gamma} \left(-m-i \omega
   +\frac{1}{2}\right)}{\mathbf{\Gamma} \left(\frac{1}{2}-m\right) \mathbf{\Gamma} \left(m-i \omega
   +\frac{1}{2}\right)}.
   \end{equation}
   This exactly takes the form of the conformal Green's function in $0+1$ dimension with a scaling dimension $\Delta=1/2-m$, which matches eq.~\eqref{SYKBH} for $m=1/4$, if we also make the substitution $\omega \to \omega \beta/2\pi$ due to the relation between boundary and global Rindler time $t=\beta t_R/2\pi$.

	Similarly, one can consider the wormhole geometry, where there are two boundaries. The solution of Dirac equations can also be derived analytically. Its asymptotic form near each boundary also contains two terms with different powers of coordinate as in eq.~\eqref{rinfBH}, whose coefficients determine the value of the source term and the operator expectation. Similar calculations on a global AdS can reproduce eq.~\eqref{Wormhole}.

	\subsection{Wormhole Geometry}
	Having introduced the basic idea of holographic calculations, we are now ready to explain the calculation of the tunneling probability. We first focus on the wormhole phase where the correlation between the left and the right system is from a non-trivial bulk geometry. The MQ model in the wormhole phase is equivalent to a fermionic field in the global AdS$_2$ spacetime with mass $m=1/2-\Delta$. Recall that we assume the leads are described by massless Dirac fermions on flat spacetime, we consider the following action of bulk Dirac fermions on the geometry shown in Fig.~\ref{fig:holographic}(a)
	\begin{equation}
	S_{\text{bulk}}^{\text{WH}}=\int \sqrt{-g}dxdt_g\ \left(i\overline{\Psi}\slashed{D}\Psi - m(x)\overline{\Psi}\Psi\right),
	\end{equation}
	where $\slashed{D}$ is the covariant derivative of spinors on curved spacetime. The metric $ds^2=(-dt_g^2+dx^2)/\zeta^2(x)$ and the mass $m(x)$ depend on the spatial coordinate as
	\begin{equation}
	(\zeta(x),m(x))=\left\{
	\begin{aligned}
	& (\epsilon,0)\ \ \ \ \ \ x\in(-\infty,\epsilon)\cup(\pi-\epsilon,\infty),& \\
	& (\sin x,1/4)\ \ \ \ \ \ x\in(\epsilon,\pi-\epsilon).&
	\end{aligned}
	\right.
	\end{equation}
	Here $\epsilon$ is a cutoff where we glue different geometries and it should be related to $\Gamma$. We also assume that $\epsilon \ll 1$ and neglected higher orders of $\epsilon$. As we will see, this corresponds to a large coupling $\lambda$ in the original model. Note that we have assumed a solid background of AdS$_2$ geometry for the MQ model, which is a direct consequence of having small number of probe fields. When the external field contains $O(N)$ degree of freedom, there will be back-reaction which changes the AdS$_2$ geometry.
	
	The tunneling probability $|T(\omega)|^2$ corresponds to a scattering problem in the gravity theory. Writing the Dirac field as $\Psi=(\psi_-,\psi_+)$, in the $x<\epsilon$ region, we consider an ingoing right-moving fermion $\psi_+(x)= e^{i\omega x}$ with momentum $p=\omega$, and a reflected left-moving fermion $\psi_-(x)= R(\omega)e^{-i\omega x}$. In the $x>\pi-\epsilon$ region, only right-moving component $\psi_+(x)=T_{\text{bulk}}(\omega)e^{i\omega x}$ exists. 

	To determine $T_{\text{bulk}}(\omega)$, we need to match the boundary condition between the global AdS spacetime and the flat spacetime by requiring the Dirac field is smooth. The complete solution of the Dirac equation in this coordinate is given in Appendix~\ref{appendix}. In the limit of small $\epsilon$, we only need the asymptotic behavior of $\Psi$ near $x=0$ and $x=\pi$. Near $x=0$, we have the expansion for the spinor fields
	\begin{equation}
    \begin{aligned}
    &\psi_{\mathbf{1}}^L(x) =  2^{-1/4}  \tilde{C_1}  x^{1/4},\\
    &\psi_{\mathbf{2}}^L(x) = 2^{1/4} \tilde{C_2}  x^{-1/4},
    \end{aligned}
	\end{equation}
	while expanding around $x=\pi$, we find
	\begin{equation}
    \begin{aligned}
    &\psi_{\mathbf{1}}^R(x) = \frac{2^{3/4} \pi}{(\pi -x)^{1/4}}  \left(\frac{\tilde{C_1} \mathbf{\Gamma} \left(\frac{3}{4}\right)}{\mathbf{\Gamma}
		\left(\frac{1}{4}\right)  \mathcal{D}_{3/4}(\omega)  }+\frac{i \tilde{C_2} \omega }{ \mathcal{D}_{1}(\omega) }\right),\\
	&\psi_{\mathbf{2}}^R(x) = 2^{1/4} \pi  (\pi -x)^{1/4} \left(\frac{\tilde{C_2} \mathbf{\Gamma}
		 	\left(\frac{1}{4}\right)}{\mathbf{\Gamma} \left(\frac{3}{4}\right) \mathcal{D}_{1/4}(\omega)}+\frac{i
		 	\tilde{C_1} \omega }{ \mathcal{D}_1(\omega) }\right).
    \end{aligned}
	\end{equation}
	Here $L/R$ represents that the general solutions with constants $\tilde{C_1}$ and $\tilde{C_2}$ are expanded around $x=0$ or $x=\pi$. As we define in Appendix~\ref{appendix}, the linear combinations of these spinor fields give rise to the left-moving or right-moving component at each boundary. The continuous condition of the left-moving component at the right boundary $x=\pi - \epsilon$ leads to the constraint $\psi_-^R(\pi-\epsilon)=(\psi_{\mathbf{1}}^R(\pi-\epsilon)-\psi_{\mathbf{2}}^R(\pi-\epsilon))/(2\iu) =0$, which finally can be simplified as
	\begin{equation}\label{eq:S_WH_rightBoundaryCond}
		\frac{\tilde{C_1}}{\tilde{C_2}}= \frac{\sqrt{2} \left(  \mathcal{D}_{3/4}(0) \cos (2 \pi  \omega ) \mathcal{D}_{1/4}(\omega) -2 \iu \pi ^2
			\sqrt{\epsilon } \sin (\pi  \omega )\right)}{\sqrt{\epsilon } \mathcal{D}_{1/4}(0) \cos (2 \pi  \omega ) \mathcal{D}_{3/4}(\omega)-4 \iu \pi ^2 \sin (\pi  \omega )}.
	\end{equation}
	Moreover, the continuous conditions of the right-moving component at the left or right boundary give that 
	\begin{equation}
	\begin{aligned}
	&e^{i\omega\epsilon}=\left( \psi_{\mathbf{1}}^L(\epsilon)+\psi_{\mathbf{2}}^L(\epsilon) \right)/2,\\&T_{\text{bulk}}(\omega)e^{i\omega(\pi-\epsilon)}=\left( \psi_{\mathbf{1}}^R(\pi-\epsilon)+\psi_{\mathbf{2}}^R(\pi-\epsilon)  \right)/2.
	\end{aligned}
	\end{equation}
	These two equations determine the tunneling probability $|T_{\text{bulk}}(\omega)|^2$ as
	\begin{equation}\label{eq:S_WH_tunneling_rates}
		\begin{aligned}
				|T_{\text{bulk}}(\omega)|^2 
				&=  \left| e^{\iu \omega (2\epsilon-\pi)} \frac{\psi_{\mathbf{1}}^R(\pi-\epsilon)+\psi_{\mathbf{2}}^R(\pi-\epsilon)}{\psi_{\mathbf{1}}^L(\epsilon)+\psi_{\mathbf{2}}^L(\epsilon)}  \right|^2 \\
				&=  \frac{8}{\frac{4 \mathcal{D}_{3/4}(0)^2}{\epsilon 
						 \mathcal{D}_{3/4}(\omega)^2 }+\frac{\epsilon  \cos ^2(2 \pi  \omega ) \mathcal{D}_{3/4}(\omega)^2 }{\mathcal{D}_{3/4}(0)^2}-4 \cos (2 \pi  \omega )+8}. \\
		\end{aligned}
	\end{equation}
    Finally, we find  $|T_{\text{bulk}}(\omega)|^2$ becomes exactly the same as eq.~\eqref{wormhole res} once we identify $\epsilon=Jt'\mathcal{D}_{3/4}(0)^2/\pi ^{5/2} \Gamma ^2$ and recall that in the bulk calculation we are measuring energy with respect to $t_g$, which differs from boundary energy $t$ by a factor of $t'$, i.e. $t_g=t't$ and the $\omega$ here corresponding to $\tilde{\omega}$ in eq.~\eqref{wormhole res}

	\subsection{Black Hole Geometry}	
	We then consider the bulk calculation in the black hole phase. Geometrically, the left and the right system is decoupled and each side of the MQ model can be replaced by a massive Dirac fermion in the Rindler spacetime. The bulk action then reads 
	\begin{equation}
	S_{\text{bulk}}^{\text{BH}}=\sum_{\alpha=L/R} \int \sqrt{-g}drdt_R\ \left(i\overline{\Psi}_\alpha\slashed{D}\Psi_\alpha - m(r)\overline{\Psi}_\alpha\Psi_\alpha\right),
	\end{equation}
	with $ds^2=-dt_R^2/\zeta^2(r)+\zeta^2(r)dr^2$ and $m(r)$ being 
	\begin{equation}\label{eq:BH_metric}
	(\zeta(r),m(r))=\left\{
	\begin{aligned}
	& (\tilde{\epsilon},0)\ \ \ \ \ \ r\in(1/\tilde{\epsilon},\infty),& \\
	& \left(1/\sqrt{r(r-2)},1/4\right)\ \ \ \ \ \ r\in(2,1/\tilde{\epsilon}).&
	\end{aligned}
	\right.
	\end{equation}
	Here boundary $\RN{1}: r=2$ is the location of the horizon, and the Rindler spacetime is connected to the flat spacetime at boundary $\RN{2}: r=1/\tilde{\epsilon}$, where $\tilde{\epsilon}$ is not necessarily the same as $\epsilon$. The boundary time $t$ is related to $t_R$ as $t=\beta t_R/2\pi$. Furthermore, to distinguish the left ($L$) and right ($R$) system, we introduce additional labels $\alpha=L/R$ on spinor fields and constants. Eq.~\eqref{eq:S_DiracBH_solution_req2} and eq.~\eqref{rinfBH} now read that
	\begin{equation}\label{eq:BH_field_C1}
\begin{aligned}
\psi_-^{\RN{1},\alpha}(r) =&(C_1^{\alpha}+C_2^{\alpha}) \frac{  2^{-\frac{1}{4}+\frac{i \omega }{2}} \pi  (r-2)^{-\frac{1}{4}-\frac{i
			\omega }{2}} \csc \left(\pi  \left(\frac{1}{2}+i \omega \right)\right)}{\mathbf{\Gamma} \left(\frac{1}{2}-i
	\omega \right) \mathbf{\Gamma} \left(i \omega +\frac{1}{2}\right)},  \\
\psi_+^{\RN{1},\alpha}(r) =& 
\frac{  2^{-\frac{1}{4}-\frac{i \omega }{2}} \pi
	(r-2)^{-\frac{1}{4}+\frac{i \omega}{2} } \csc \left(\pi  \left(\frac{1}{2}-i \omega \right)\right) }{ \mathbf{\Gamma} \left(i
	\omega +\frac{1}{2}\right)^2} \\
&\ \ \times\left(C_1^{\alpha} \frac{\mathbf{\Gamma}
	\left(i \omega +\frac{1}{2}+m\right)}{\mathbf{\Gamma} \left(\frac{1}{2}+m-i \omega \right)}-C_2^{\alpha}\frac{\mathbf{\Gamma}
	\left(i \omega +\frac{1}{2}-m\right)}{\mathbf{\Gamma} \left(\frac{1}{2}-m-i \omega \right)}\right).
\end{aligned}
\end{equation}
	
	\begin{equation}\label{eq:BH_field_C2}
\begin{aligned}
\begin{pmatrix}
\psi_-^{\RN{2},\alpha}\\\psi_+^{\RN{2},\alpha}
\end{pmatrix}=&C_2^{\alpha}\frac{r^{-\frac{1}{2}+m}\mathbf{\Gamma}(1-m)\mathbf{\Gamma}(\frac{1}{2}-m+i\omega)}{2^m\mathbf{\Gamma}(1-2m)\mathbf{\Gamma}(\frac{1}{2}+i\omega)} \begin{pmatrix}
1\\-1
\end{pmatrix}\\
+&C_1^{\alpha}\frac{r^{-\frac{1}{2}-m}\mathbf{\Gamma}(1+m)\mathbf{\Gamma}(\frac{1}{2}+m+i\omega)}{2^{-m}\mathbf{\Gamma}(1+2m)\mathbf{\Gamma}(\frac{1}{2}+i\omega)} \begin{pmatrix}
1\\1
\end{pmatrix}.
\end{aligned}
	\end{equation}
	
	Up to now, the left and right copy are still decoupled. To have a non-vanishing contribution, we further need to add the coupling between the boundary $\RN{2}$ in different copies
	\begin{equation}\label{cou}
	\Delta S_{\text{bulk}}^{\text{BH}}\sim \nu\int \sqrt{-\gamma} dt_R \left(\overline{\Psi^B_{L}}\Psi^B_{R}+\text{H.C.}\right)
	\end{equation}
	Here $\Psi^B_{\alpha}$ is the corresponding boundary operator at side $\alpha$ and $\gamma$ is the reduced metric on the boundary. Coupling constant $\nu$ should be proportional to $\mu$. This geometry is illustrated in Fig.~\ref{fig:holographic}(b).
	
	The tunneling probability can now be determined perturbatively. When $\nu=0$, by again imposing the in-falling boundary condition near the horizon, we can compute an ingoing Dirac fermion in the left lead scattered by the left black hole. The metric in flat space, defined in eq.~\eqref{eq:BH_metric}, reveals that the ingoing Dirac fermion is a mode $e^{-i\omega \tilde{\epsilon}^2 r}$ in the left lead which moves towards $r=2$. Note that the sign of the ingoing mode is determined by the convention of coordinates. Here the ingoing mode moves to the direction that $r$ decreases, in contrast to the wormhole phase where the ingoing mode moves to the direction that $x$ increases, as denoted in Fig.~\ref{fig:holographic}(a) and (b). The matching between the mode and eq.\eqref{rinfBH} at $r=1/\tilde{\epsilon}$ gives 
	\begin{equation}\label{eq:BH_L_Continuous}
e^{-\iu \omega \tilde{\epsilon}}= \psi_-^{\RN{2}, L}(1/\tilde{\epsilon}).
	\end{equation}
	Here we have added the $L/R$ indices for the bulk fields to distinguish two Rindler spacetimes. Together with the result of in-falling boundary conditions

		\begin{equation}\label{eq:S_DiracBH_solution_C1divC2_alpha}
\begin{aligned}
\frac{C_1^{\alpha}}{C_2^{\alpha}} = \frac{\mathbf{\Gamma} \left(\frac{1}{2}+m-i \omega \right) \mathbf{\Gamma} \left(i \omega +\frac{1}{2}-m\right)}{\mathbf{\Gamma}
	\left(\frac{1}{2}-m-i \omega \right) \mathbf{\Gamma} \left(i \omega +\frac{1}{2}+m\right)},
\end{aligned}
\end{equation}
	eq.~\eqref{eq:BH_L_Continuous} determines the wavefunction $(\psi_-^{ L}(r),\psi_+^{ L}(r))$ on the left copy. 

	We then take the coupling eq.~\eqref{cou} into account perturbatively. The coupling term can be understood as a source term on the right boundary, whose strength is proportional to $\nu$ times the expectation of the left system boundary operator \cite{hartnoll2018holographic}, i.e. the coefficient in eq.~\eqref{eq:BH_field_C2} that is proportional to $r^{-1/2+m}$. This gives 
    	\begin{equation}\label{eq:S_WH_sourceL}
		\delta J^{R} = \nu\frac{C_2^{L} \mathbf{\Gamma} \left(\frac{3}{4}\right) \mathbf{\Gamma} \left(i \omega
			+\frac{1}{4}\right)}{2^{1/4} \sqrt{\pi } \mathbf{\Gamma} \left(i \omega
			+\frac{1}{2}\right)}.
	\end{equation}
	With this additional contribution, we can determine two continuous conditions at the right copy to the leading order of $\nu$. First, the absence of the ingoing wave at boundary $\RN{2}$ of the right copy ensures that $(1/\tilde{\epsilon})^{-3/4} \delta J^{R} + \psi_{-}^{\RN{2}, R}(1/\tilde{\epsilon})=0$. This condition could be written explicitly: 
		 \begin{equation}\label{eq:S_DiracBH_Bound_cond2_in}
			\frac{2^{3/2}  \mathbf{\Gamma}(\frac{5}{4})  \mathbf{\Gamma}(i \omega+\frac{3}{4}) }{\mathbf{\Gamma}(\frac{3}{4})  \mathbf{\Gamma}(i \omega+\frac{1}{4})} C_1^R+ (\tilde{\epsilon})^{-1/2} C_2^R - \nu C_2^L=0.
	 \end{equation}
    Second, the outgoing wave can be similarly calculated as $T_{\text{bulk}}(\omega) e^{i\omega\tilde{\epsilon}} = (1/\tilde{\epsilon})^{-3/4} \delta J^{R} + \psi_{+}^{\RN{2}, R}(1/\tilde{\epsilon}) $. Finally we could obtain the tunneling probability
	 \begin{equation}\label{eq:S_DiracBH_Tbulk2}
	 	\begin{aligned}
	 		|T_{\text{bulk}}|^2 
	 		&= \left|  \frac{ (1/\tilde{\epsilon})^{-3/4} \delta J^{R} + \psi_{+}^{\RN{2}, R}(1/\tilde{\epsilon})}{\psi_-^{\RN{2}, L}(1/\tilde{\epsilon}) } \right|^2   \\
	 		&= \frac{4 \nu ^2 \tilde{\epsilon }}{\left|1+\frac{\sqrt{\tilde{\epsilon }} \mathbf{\Gamma} \left(\frac{1}{4}\right) \mathbf{\Gamma}
	 				\left(\frac{3}{4}-i \omega \right)}{\sqrt{2} \mathbf{\Gamma} \left(\frac{3}{4}\right)
	 				 \mathbf{\Gamma} \left(\frac{1}{4}-i \omega \right)}\right|^4
	 			} \\
	 	\end{aligned}
	 \end{equation}
    By identifying 
    \begin{equation}
\nu=\mu/\left(\Gamma\sqrt{\tilde{\epsilon}} \right),\ \ \ \ \  \tilde{\epsilon}={4 \sqrt{\pi } J \mathbf{\Gamma} \left(\frac{3}{4}\right)^2}/ \left( \beta  \Gamma ^2 \mathbf{\Gamma} \left(\frac{1}{4}\right)^2 \right),
    \end{equation}
    and again making the substitution $\omega \to \omega \beta/2\pi$,
    we find $|T_{\text{bulk}}(\omega)|^2$ matches the result (eq.~\eqref{BH res}) on the quantum side exactly.

	\section{Conclusion} 
	We consider the tunneling spectroscopy for the MQ model by coupling each side to a different lead. In the low-temperature wormhole phase and for small coupling to leads, both the tunneling probability $|T(\omega)|^2$ and the differential conductance $dJ_R(V)/dV$ show peaks at $|\omega_n|=t'(1/4+n)$, which is fixed by the $\SL(2)$ symmetry. In the high-temperature black hole phase, there is only a single peak near $\omega=0$. We further give a holographic picture in both phases and find an exact match for the calculation between the gravity side and the quantum side. 
	
	There are several extensions of the current work. One can consider adding chemical potential to the complex version of the MQ model, which should be dual to adding gauge fields in the bulk. It is also interesting to consider a large number of modes in the leads, and then there will be non-trivial back-reaction for the AdS$_2$ background and the problem should be solved self-consistently. We defer these to further studies.

	\section{Acknowledgment}
	We thank Yingfei Gu for helpful discussions. PZ also thanks Yiming Chen and Xiao-Liang Qi for valuable discussions when collaborating on previous works \cite{Chen:2020wiq,Chen:2019qqe}. PZ acknowledges support from the Walter Burke Institute for Theoretical Physics at Caltech.
    
    \appendix
    \begin{widetext}

    \section{The Solution of the Dirac Equation}\label{appendix}
    In this appendix, we give the general solutions of the Dirac equation in both the wormhole geometry and the blackhole geometry.

    We firstly consider the global AdS$_2$ spacetime for the wormhole geometry with metric $ds^2=(-dt_g^2+dx^2)/\sin^2 x$. In terms of left-moving and right-moving component $\Psi = \sqrt{\sin x}(\psi_-, \psi_+)$, the Dirac equation reads 
			\begin{equation}\label{eq:S_DiracWH_1}
	\begin{aligned}
	& \iu (\partial_{t_g} + \partial_{x}) \psi_+(t_g, x) =-m \psi_-(t_g, x) \csc (x) ,  &  \\
	& \iu (\partial_{t_g} - \partial_{x}) \psi_-(t_g, x) =-m \psi_+(t_g, x) \csc (x) . &
	\end{aligned}
	\end{equation}
	Specifically, left-moving is defined as the direction $x$ decreases and vice versa. After performing Fourier transform on the global time $t_g$ and get the corresponding frequency $\omega$, the equation becomes
			\begin{equation}\label{eq:S_DiracWH_2}
	\begin{aligned}
	& \omega  \psi_+(\omega, x) + \iu \partial_{x}\psi_+(\omega, x)=-m \psi_-(\omega, x) \csc (x) ,  &  \\
	& \omega  \psi_-(\omega, x)  - \iu \partial_{x}\psi_-(\omega, x)=-m \psi_+(\omega, x) \csc (x) . &
	\end{aligned}
	\end{equation}
	Next we will abbreviate $ \psi_{-/+}(\omega, x)$ as $\psi_{-/+}(x)$ for simplicity. This set of differential equations can be analytically solved. By introducing new variable $\psi_{\mathbf{1}}(x) = \psi_+(x)+\iu \psi_-(x), \psi_{\mathbf{2}}(x) = \psi_+(x)-\iu \psi_-(x)$ (here we use bold index $\mathbf{1}$ and $\mathbf{2}$ to avoid possible confusion with the label on the Keldysh contour), the solution of \eqref{eq:S_DiracWH_2} reads \footnote{We acknowledge the the solutions of \eqref{eq:S_DiracWH_2} are obtained by Yiming Chen.}
	\begin{equation}\label{eq:S_DiracWH_solution}
	\begin{split}
			\psi_{\mathbf{1}}(x)	&= \frac{(1+\cos (x) )^{\frac{1}{4}-\frac{m}{2}}} {\sqrt{\sin (x)}} \Bigg[\tilde{C_1} (1-\cos
	(x))^{\frac{1}{4}+\frac{m}{2}} \, _2F_1\left(-\omega ,\omega ;m+\frac{1}{2};\sin
	^2\left(\frac{x}{2}\right)\right) \\
	& \qquad\qquad +\tilde{C_2}\frac{i 2^{\frac{1}{2}+m} \omega }{1-2m} (1-\cos
	(x))^{\frac{3}{4}-\frac{m}{2}} \, _2F_1\left(-m-\omega +\frac{1}{2},-m+\omega
	+\frac{1}{2};\frac{3}{2}-m;\sin ^2\left(\frac{x}{2}\right)\right) \Bigg], \\
				\psi_{\mathbf{2}}(x)	&= \frac{(1+\cos (x) )^{\frac{1}{4}+\frac{m}{2}}} {\sqrt{\sin (x)}} \Bigg[\tilde{C_1} (1-\cos
	(x))^{\frac{1}{4}-\frac{m}{2}} \, _2F_1\left(-\omega ,\omega ;\frac{1}{2}-m;\sin
	^2\left(\frac{x}{2}\right)\right) \\
	& \qquad\qquad +\tilde{C_2}\frac{i 2^{\frac{1}{2}-m} \omega }{1+2m} (1-\cos
	(x))^{\frac{3}{4}+\frac{m}{2}} \, _2F_1\left(\frac{1}{2} +m-\omega ,\frac{1}{2} +m+\omega;\frac{3}{2}+m;\sin ^2\left(\frac{x}{2}\right)\right) \Bigg], \\
	\end{split}
	\end{equation}
	where $\, _2F_1(a,b;c;y)$ is the standard hypergeometric function and we have two undetermined constants $\tilde{C}_1$ and $\tilde{C}_2$.

	We then consider the black hole geometry with metric $ds^2=-r(r-2)dt_R^2+dr^2/r(r-2)$. Now we define the $\Psi = (\psi_-, \psi_+)$ which gives 
		\begin{equation}\label{eq:S_DiracBH_1}
		\begin{aligned}
		 -\frac{\partial_{t_R} \psi_-(t_R, r)}{\sqrt{r(r-2)}} + \sqrt{r(r-2)}\partial_{r} \psi_-(t_R, r)+\frac{r-1}{2\sqrt{r(r-2)}} \psi_- (t_R, r) + m \psi_+(t_R, r) =0,  &  \\ 
 		 \frac{\partial_{t_R} \psi_+(t_R, r)}{\sqrt{r(r-2)}} + \sqrt{r(r-2)}\partial_{r} \psi_+(t_R, r)+\frac{r-1}{2\sqrt{r(r-2)}} \psi_+(t_R, r) + m \psi_-(t_R, r)=0 ,  &  \\
		\end{aligned}
		\end{equation}
		The bulk wave function moving to the direction that $r$ increases when $m=0$ is labeled by $\psi_+(t_R, r)$, and vice versa. Then we perform Fourier transform on the Rindler time $t_R$ and get the corresponding frequency $\omega$. The equation becomes
	\begin{equation}\label{eq:S_DiracBH_2}
	\begin{aligned}
	\frac{\iu \omega \psi_-(r)}{\sqrt{r(r-2)}} + \sqrt{r(r-2)}\partial_{r} \psi_-(r)+\frac{r-1}{2\sqrt{r(r-2)}} \psi_-(r) + m \psi_+(r) =0.  &  \\ 
	-\frac{\iu \omega \psi_+(r)}{\sqrt{r(r-2)}} + \sqrt{r(r-2)}\partial_{r} \psi_+(r)+\frac{r-1}{2\sqrt{r(r-2)}} \psi_+(r) + m \psi_-(r)=0 ,  &  \\
	\end{aligned}
	\end{equation}
	Here we also abbreviate $ \psi_{-/+}(\omega, r)$ as $\psi_{-/+}(r)$. The solutions of these differential equations have the form \cite{gu2020notes}
	\begin{equation}\label{eq:S_DiracBH_full_solution}
		\begin{aligned}
				\psi_-(r) =\   & \Bigg( C_1 \frac{ 2^m \left(\frac{1}{r}\right)^m \mathbf{\Gamma} (m+1) \left(1-\frac{2}{r}\right)^{-\frac{i
						\omega }{2}} \mathbf{\Gamma} \left(m+i \omega +\frac{1}{2}\right) \, _2F_1\left(m,m-i \omega +\frac{1}{2};2
				m+1;\frac{2}{r}\right)}{(r^2-2 r)^{1/4} \mathbf{\Gamma} (2 m+1) \mathbf{\Gamma} \left(i \omega
				 +\frac{1}{2}\right)}  \\
			 & +C_2 \frac{ 2^{-m} \left(\frac{1}{r}\right)^{-m} \mathbf{\Gamma} (1-m)
				\left(1-\frac{2}{r}\right)^{-\frac{i \omega }{2}} \mathbf{\Gamma} \left(-m+i \omega +\frac{1}{2}\right) \,
				_2F_1\left(-m,-m-i \omega +\frac{1}{2};1-2 m;\frac{2}{r}\right)}{(r^2-2 r)^{1/4} \mathbf{\Gamma} (1-2 m)
				\mathbf{\Gamma} \left(i \omega +\frac{1}{2}\right)} \Bigg) \\
							\psi_+(r) =\   & \Bigg( C_1\frac{ 2^m \left(\frac{1}{r}\right)^m \mathbf{\Gamma} (m+1) \left(1-\frac{2}{r}\right)^{\frac{i
						\omega }{2}} \mathbf{\Gamma} \left(m+i \omega +\frac{1}{2}\right) \, _2F_1\left(m,m+i \omega +\frac{1}{2};2
				m+1;\frac{2}{r}\right)}{(r^2-2 r)^{1/4} \mathbf{\Gamma} (2 m+1) \mathbf{\Gamma} \left(i \omega
				+\frac{1}{2}\right)} \\
			&- C_2 \frac{ 2^{-m} \left(\frac{1}{r}\right)^{-m} \mathbf{\Gamma} (1-m)
				\left(1-\frac{2}{r}\right)^{\frac{i \omega }{2}} \mathbf{\Gamma} \left(-m+i \omega +\frac{1}{2}\right) \,
				_2F_1\left(-m,-m+i \omega +\frac{1}{2};1-2 m;\frac{2}{r}\right)}{(r^2-2 r)^{1/4} \mathbf{\Gamma} (1-2 m)
				\mathbf{\Gamma} \left(i \omega +\frac{1}{2}\right)} \Bigg) \\
		\end{aligned}
	\end{equation}
	Here we have two undetermined constants $C_1$ and $C_2$.
 \end{widetext}

\end{document}